\documentclass[aps,reprint,amsmath,amssymb,prb,superscriptaddress,floatfix]{revtex4-1}
\usepackage[pdftex]{graphicx}
\usepackage{dcolumn}
\usepackage{bm}
\usepackage{braket}
\usepackage{color}

  \PassOptionsToPackage{monochrome}{xcolor}
\graphicspath{{img-gray/}}
\newcommand{\kv}{\boldsymbol{k}}
\newcommand{\qv}{\boldsymbol{q}}
\allowdisplaybreaks[4] 

\begin{document}


\title{Improved tetrahedron method for the Brillouin-zone integration applicable to response functions}



\author{Mitsuaki Kawamura}
\email{kawamura@cms.phys.s.u-tokyo.ac.jp}

\author{Yoshihiro Gohda}
\affiliation{Department of Physics, The University of Tokyo, Tokyo 113-0033, Japan}

\author{Shinji Tsuneyuki$^{1,}$}
\affiliation{Institute for Solid State Physics, 
  The University of Tokyo, Kashiwa 277-8581, Japan}



\date{\today}

\begin{abstract}
  We improve the linear tetrahedron method to overcome systematic errors
  due to overestimations (underestimations) in integrals for 
  convex (concave) functions, respectively.  
  Our method is applicable to various types of calculations such as the total
  energy, the charge (spin) density, response functions, and the
  phonon frequency, in contrast with the Bl\"ochl correction, which is
  applicable to only the first two.
  We demonstrate the ability of
  our method by calculating phonons in MgB$_2$ and fcc lithium.
\end{abstract}


\maketitle
%
%
\section{Introduction}
In calculations of periodic systems on the basis of density functional
theory (DFT)\cite{PhysRev.136.B864}, integrals of matrix elements over
the Brillouin zone (BZ) are evaluated to obtain various physical quantities
of solids including the total energy, the electron (spin)
density, the density of states, response functions, and the phonon
frequency.  Since this integral with respect to the Bloch wave vector
$\kv$ is replaced with a summation over a range of points described by a discrete variable $\kv$,
approximation schemes employed for this summation can significantly
affect the accuracy and computational costs.
Accurate integration using a modest number of $\kv$ points is even more important for 
hybrid-DFT \cite{ISI:A1996VW94700028} and 
$GW$ approximation \cite{PhysRev.139.A796}, 
because in these cases the computational cost is proportional to the square of the number of $\kv$ points,
whereas standard semi-local approximations have a linear dependence.

There are two kinds of schemes to perform such an integration over the $\kv$ points,
namely, 
the broadening method\cite{PhysRevB.40.3616} and 
the tetrahedron method\cite{Jepson19711763}.
In the broadening method,
we replace the delta function with a smeared function
which has a finite broadening width;
we have to check the convergences about both the broadening width and 
the number of $\kv$ points to obtain accurate results. 
In the tetrahedron method,
we perform analytical integration in tetrahedral regions covering the BZ
with the piecewise linear interpolation of a matrix element.
Unlike the broadening method,
we have to check the convergence only about the number of $\kv$ points.
The tetrahedron method is applied to calculations of 
susceptibility \cite{PhysRevB.11.2109},
phonon frequency \cite{PhysRevLett.69.2819},
phonon line width \cite{PhysRevB.54.16487}, and
the local Green's function as part of the dynamical mean field theory
in the Hubbard model \cite{JPSJ.72.777}.

However, the tetrahedron method has a drawback;
if a matrix element $A_{\kv}$ is a convex (concave) function of $\kv$,
this method systematically overestimates (underestimates) its contribution to the integral 
due to the linear interpolation involved.
Although this can be avoided by using the quadratic interpolation,
we cannot perform analytical integration straightforwardly in such a case.
The Bl\"ochl correction \cite{PhysRevB.49.16223} was invented to overcome this issue
by utilizing the following two facts:
(i) 
the difference between the linear interpolation integration
and that using the quadratic interpolation 
is approximately proportional to 
the second derivative of $A_{\kv}$ integrated over the occupied region;
(ii) although $\partial^2 A_{\kv}/\partial k^2$ cannot be evaluated 
within the framework of linear interpolation,
we can perform the volume integral by replacing it with 
the Fermi surface integration of the first derivative of $A_{\kv}$
(which can be evaluated by linear interpolation) 
using the Gau{\ss} theorem.
Using this method we can reduce the number of $\kv$ points to obtain converged results for
total energies and charge densities.
However, 
in the calculation of response functions or phonon frequencies,
the integral $\int_{\epsilon_{\kv} < \epsilon_{\rm F}} d^3 k A_{\kv}/(\epsilon_{\rm F} - \epsilon_{\kv})$
appears, where $A_{\kv}$ is an arbitrary function of $\kv$.
In this case, 
the Bl\"ochl correction is inapplicable
because we cannot perform the  Fermi surface integration 
when $\partial (A_{\kv}/(\epsilon_{\rm F} - \epsilon_{\kv}))/\partial k$ is singular.

In this work, 
we develop a newly improved tetrahedron method
that is applicable to calculations involving integrations of functions
with singularities on the Fermi surfaces.
It is constructed by means of a higher-order interpolation and the least square method.
We apply our method to the BZ integration in calculations of
phonon frequencies based on density functional perturbation theory (DFPT)
\cite{RevModPhys.73.515}.
Following that we successfully calculate the frequency of phonons in MgB$_2$ and fcc Li.
In contrast, it is difficult to achieve convergence in this calculation using conventional methods
because the phonons in these materials couple strongly with electrons
in the vicinity of Fermi surfaces
\cite{PhysRevB.82.165111, PhysRevB.82.184509}.
In Sec. II, 
we describe our new tetrahedron method in detail after summarizing the conventional
linear tetrahedron method and 
the Bl\"ochl correction.
Section III shows how our method improves the convergence about the number of $\kv$ points 
in the calculation of phonons, followed by the conclusion in Sec. IV.

%
%
\section{Method}
In this section, we introduce our new tetrahedron method;
we begin with the standard linear tetrahedron method and the Bl\"ochl correction
to explain why these methods are not necessarily efficient in calculating response 
functions such as phonon frequencies.
\subsection{The linear tetrahedron method and its drawbacks}
%
%
\begin{figure}[!tb]
  \includegraphics[width=8.5cm]{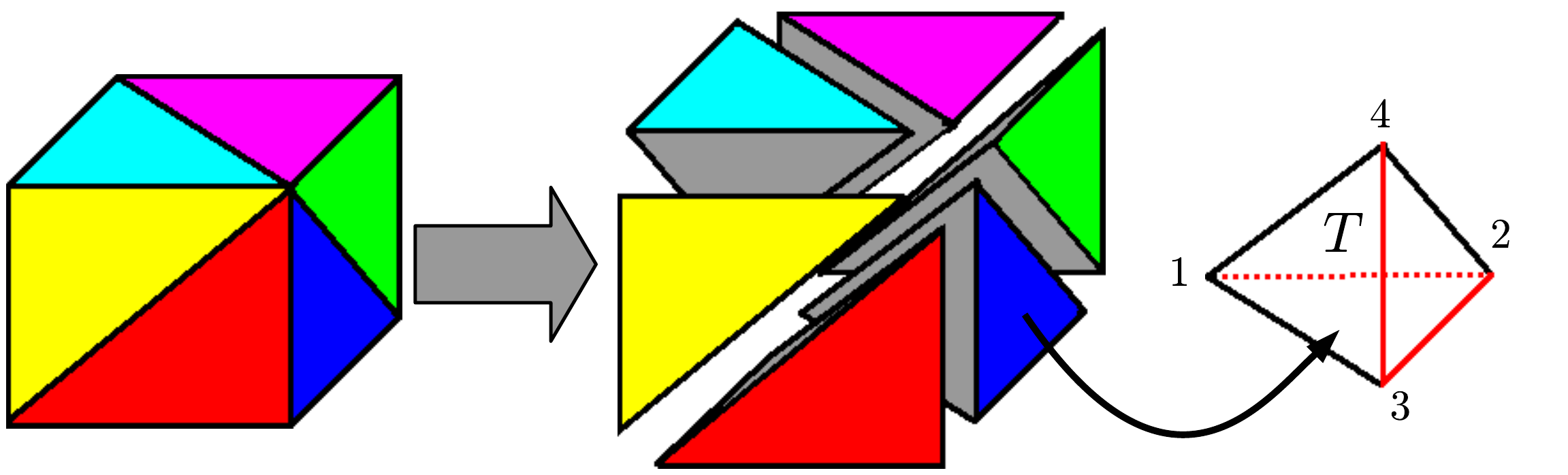}
  \caption{\label{cube_to_tetra}Sub-cell division into six tetrahedra
    and numbering of the tetrahedron corners;
    the red lines in the rightmost tetrahedron are the edges of the sub-cell.}
\end{figure}

We overview the general procedure of the tetrahedron method and its drawbacks.
We calculate the integral
\begin{align}\label{k_int}
  \int d^3 k A_{\kv} B(\epsilon_{\kv}),
\end{align}
on the basis of the linear tetrahedron method,
where $B(\epsilon_{\kv})$ is a function of the orbital energy such as
$\theta(\epsilon_{\rm F} - \epsilon_{\kv})$, 
$\delta(\epsilon_{\rm F} - \epsilon_{\kv})$,
or $\theta(\epsilon_{\rm F} - \epsilon_{\kv}) / (\epsilon_{\rm F} - \epsilon_{\kv})$.
Here, $\theta(x)$ is the Heaviside step function.
First, we divide a sub-cell into six tetrahedra (Fig. \ref{cube_to_tetra});
this sub-cell is partitioned with the uniform $\kv$-point mesh;
for convenience, we number the corners of each tetrahedron
from 1 to 4 along specific edges of the sub-cell (see Fig. \ref{cube_to_tetra}).
The contribution of this tetrahedron ($T$) to the integral (\ref{k_int}) is
\begin{align}\label{tet_int}
  6 V_T \int_0^1 dx \int_0^{1-x} dy \int_0^{1-x-y} dz
  A_T({\boldsymbol s}) B[\epsilon_T ({\boldsymbol s})],
\end{align}
where ${\boldsymbol s} = (x, y, z)$, and
\begin{align}
  A_T({\boldsymbol s}) \equiv A_{\kv_1^T(1-x-y-z) 
  + \kv_2^T x + \kv_3^T y + \kv_4^T z},
  \\
  \epsilon_T({\boldsymbol s}) \equiv \epsilon_{\kv_1^T(1-x-y-z) 
  + \kv_2^T x + \kv_3^T y + \kv_4^T z },
\end{align}
where $\kv_{i}^{T}$ is the $\kv$ point of the $i$th corner of $T$.
In the linear tetrahedron method, we approximate $A_T$ and $\epsilon_T$
with linear functions:
\begin{align}
  \label{lint_f} A_T^{1}({\boldsymbol s})
  &= A_1 (1-x-y-z) + A_2 x + A_3 y + A_4 z,
  \\
  \label{lint_e} \epsilon_{T}^1 ({\boldsymbol s}) 
  &= \epsilon_1 (1-x-y-z) 
  + \epsilon_2 x + \epsilon_3 y + \epsilon_4 z,
\end{align}
where $A_i$ and $\epsilon_i$ are the matrix element and the orbital energy at the $i$th corner, respectively. 
The integration (\ref{tet_int}) 
with formulae (\ref{lint_f}) and (\ref{lint_e}) is performed {\it analytically}.

\begin{figure}[!tb]
  \includegraphics[width=8.5cm]{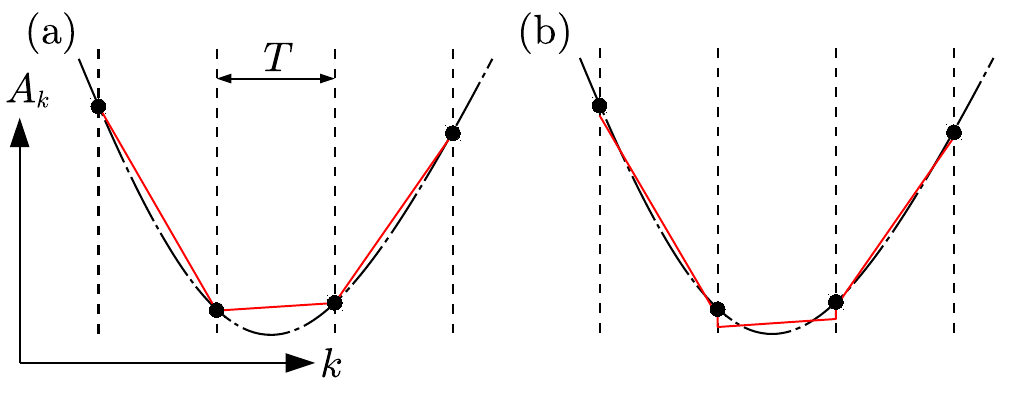}
  \caption{\label{levelfig}
   Two kinds of approximations of the matrix element.
    True and approximated matrix elements $A_T$ are depicted as black dash-doted lines 
    and red solid lines, respectively;  the black points indicate the matrix elements $A_{\kv}$
    for a given value of $\kv$; the dashed lines indicate the boundaries of the tetrahedra.
    (a) The liner interpolated function $A_T^1$ is always smaller (larger) than 
    the true function $A_T$ in the case of a convex (concave) function. 
    (b) The leveled linear function is a better approximation of the true function.}
\end{figure}

However, linear interpolation has a drawback;
if the matrix element $A_T({\boldsymbol s})$ 
is a convex function within the tetrahedron $T$,
the interpolated function $A_T^{1}({\boldsymbol s})$ becomes
$A_T^{1}({\boldsymbol s}) \ge A_T({\boldsymbol s})$
in $T$;
hence, the integral is systematically overestimated.
If $A_T({\boldsymbol s})$ is a concave function,
the sign of the inequality is reversed (see Fig. \ref{levelfig}(a)).

\subsection{The Bl\"ochl correction and its limitation}

In the special case that the integral (\ref{k_int}) becomes 
\begin{align}\label{kineng}
  \int d^3 k A_{\kv} \theta(\epsilon_{\rm F} - \epsilon_{\kv}),
\end{align}
we can overcome the drawback of the linear tetrahedron method
by considering the curvature of $A_{\kv}$
within the framework of the linear interpolation \cite{PhysRevB.49.16223};
this type of integration appears in the calculations of 
total energies or charge (spin) densities.
In this case,
we can evaluate the difference between the integral (\ref{kineng}) with the linear interpolation 
of $A_{\kv}$ ($A^{\rm lin}$) and that with the quadratic interpolation ($A^{\rm quad}$)
as follows.
First, we write this difference as
\begin{align}
  \Delta A \equiv A^{\rm quad} - A^{\rm lin} = \sum_{T}^{\epsilon_{\kv} \le \epsilon_{\rm F}} \sum_{i j} C_{i j}^{T}
  \Braket{ \frac{\partial^2 A_{\kv}}{\partial k_i \partial k_j}}_{T},
\end{align}
where $C_{i j}^{T}$ is the form factor 
describing the shape and the orientation of the tetrahedron as follows
\begin{align}
  \hspace{-0.1em}
  C_{i j}^{T} = \frac{1}{40}\left[
   \sum_{l=1}^4 (\kv_l^T)_i \sum_{m=1}^4 (\kv_m^T)_j 
   - 4 \sum_{l=1}^4 (\kv_l^T)_i (\kv_l^T)_j.
   \right],
\end{align}
and $\Braket{ \cdots}_{T}$ indicates an integration in the tetrahedron $T$.
Now, we replace $\partial^2 A_{\kv}/\partial k^2$ with $\partial A_{\kv}/\partial k$ 
because the former cannot be evaluated in the framework of the linear interpolation,
but the latter can be.
We assume the form factor is a constant over the entire BZ ($C_{i j}^T \approx C_{i j}$),
and then we apply the Gau{\ss} theorem:
\begin{align}
  \Delta A 
  & \approx \sum_{i j}  C_{i j} \int_{\epsilon_{\kv} < \epsilon_{\rm F}}
  d^3k \frac{\partial^2 A_{\kv}}{\partial k_i \partial k_j}
  \nonumber \\
  & = \sum_{i j}  C_{i j} \int_{\epsilon_{\kv} = \epsilon_{\rm F}}
  d^2 k \frac{(\nabla_{\kv} \epsilon_{\kv})_i}{|\nabla_{\kv} \epsilon_{\kv}|}
  \frac{\partial A_{\kv}}{\partial k_j}
  \nonumber \\
  &\approx \sum_{T}^{{\rm Fermi surface}} \sum_{i j } C_{i j}^{T}
  \left \langle
    \frac{(\nabla_{\kv} \epsilon_{\kv})_i}{|\nabla_{\kv} \epsilon_{\kv}|}
    \frac{\partial A_{\kv}}{\partial k_j}
  \right \rangle_T
\end{align}

However, when we calculate an integral such as
\begin{align}
  \int d^3 k A_{\kv} \frac{\theta(\epsilon_{\rm F} - \epsilon_{\kv})}{\epsilon_{\rm F} - \epsilon_{\kv}},
\end{align}
(this kind of integration appears in the calculations of response functions
and phonon frequencies),
the difference associated with the two kinds of interpolation
becomes 
\begin{align}
  \Delta A = \sum_{T}^{\epsilon_{\kv} < \epsilon_{\rm F}} \sum_{i j} C_{i j}^{T}
  \Braket{ \frac{\partial^2 A_{\kv}}{\partial k_i \partial k_j}}_{T}
  G(\epsilon_{\kv}, \nabla_{\kv} \epsilon_{\kv}),
\end{align}
where $G(\epsilon_{\kv}, \nabla_{\kv} \epsilon_{\kv})$ is a complicated function of 
$\epsilon_{\kv}$ and $\nabla_{\kv} \epsilon_{\kv}$;
therefore, we cannot apply the Bl\"ochl correction 
because we cannot replace $\partial^2 A_{\kv}/\partial k^2$ with $\partial A_{\kv}/\partial k$
as before.
This is due to the presence of the energy denominator;
hence, we have to start with another concept 
to overcome this issue.

\subsection{A newly improved tetrahedron method applicable to response functions}
The systematic error of the tetrahedron method is a result of the {\it linear interpolation}.
Although we can avoid this problem if we use higher order interpolation, the integral (\ref{tet_int}) 
becomes unsolvable analytically.
The real question is: how can we improve the {\it linear} approximation of the matrix elements?
The answer is to employ {\it leveling} rather than interpolating
(see fig. \ref{levelfig} b).
The procedure is explained below.
\begin{enumerate}
  \item We construct the $N$th polynomial $A_T^N({\boldsymbol s})$ 
    from $A_{\kv}$ and $\kv$ using the corners of a tetrahedron $T$
    and some additional surrounding points for sampling.
  \item We fit a linear function
    \begin{align}\label{eqn_ftlsm}
      \hspace{-0.3em}
      A_T^{LSM}({\boldsymbol s}) =
      {\bar A}_1 (1-x-y-z) + {\bar A}_2 x + {\bar A}_3 y + {\bar A}_4 z
    \end{align}
    into $A_T^N({\boldsymbol s})$ through the least square method (LSM);
    that is to say, we solve
    \begin{align}\label{eqn_lsm}
      \frac{\partial}{\partial {\bar A}_i} \int_0^1 dx &\int_0^{1-x} dy \int_0^{1-x-y} dz 
      \nonumber \\
      & \times |A_T^N({\boldsymbol s}) - A_T^{LSM}({\boldsymbol s})|^2 = 0.
    \end{align}
  \item We apply the same procedure to $\epsilon_{\kv}$, 
    and obtain $\epsilon_{T}^{LSM}({\boldsymbol s})$.
  \item We evaluate integral (\ref{tet_int}) replacing 
    $A_{T}({\boldsymbol s})$ and $\epsilon_{T}({\boldsymbol s})$ with
    $A_{T}^{LSM}({\boldsymbol s})$ and $\epsilon_{T}^{LSM}({\boldsymbol s})$, respectively.
  \item We repeat the above steps for all tetrahedra.
\end{enumerate}
Although the approximated matrix element $A_T^{LSM}({\boldsymbol s})$ is discontinuous
at boundaries of tetrahedra (see Fig. \ref{levelfig}(b)), 
it is of no concern 
because we are interested only in the integrated value.
\subsection{Implementation}
%
%
\begin{figure}[!tb]
  \includegraphics[width=8.5cm]{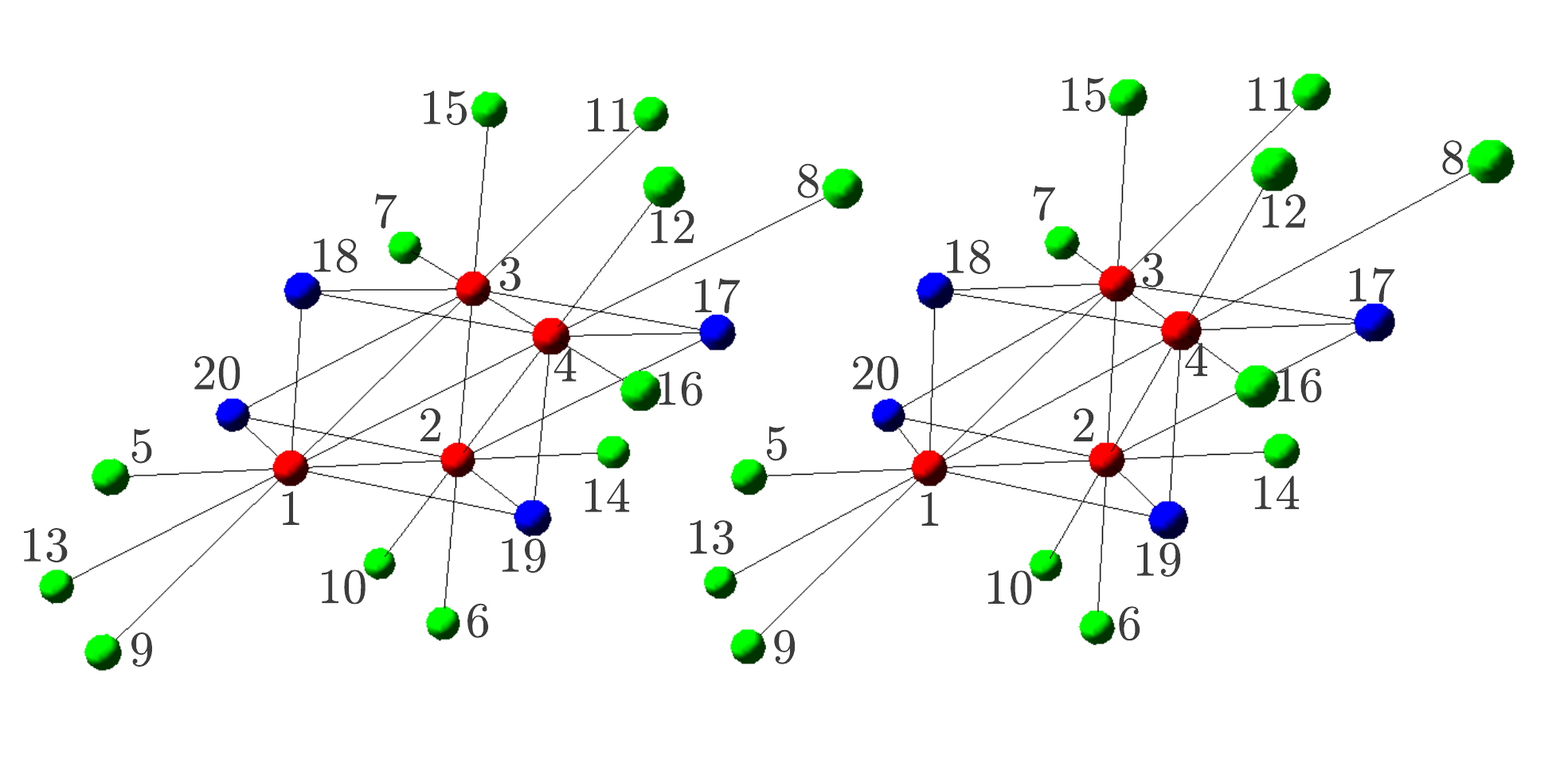}
  \caption{\label{pointsfig}
    Points for constructing a third order interpolation function
    (parallel stereogram). Red points denote the corners of $T$.
    The blue and green points are explained in Table \ref{pointstable}.}
\end{figure}
\begin{table}[!bt]
  \caption{\label{pointstable}Points for constructing a third order interpolation function}
  \begin{ruledtabular}
    \begin{tabular}{ccc}
      \multicolumn{3}{c}{
        Nearest neighbor points on extended lines }\\
       \multicolumn{3}{c}{
         of each edge of $T$ (green balls in Fig. \ref{pointsfig}).} \\
       \hline
      $\kv_{ 5} = 2 \kv_1 - \kv_2$ & $\kv_{ 9} = 2 \kv_1 - \kv_3$ & $\kv_{13} = 2 \kv_1 - \kv_4$ \\
      $\kv_{ 6} = 2 \kv_2 - \kv_3$ & $\kv_{10} = 2 \kv_2 - \kv_4$ & $\kv_{14} = 2 \kv_2 - \kv_1$ \\
      $\kv_{ 7} = 2 \kv_3 - \kv_4$ & $\kv_{11} = 2 \kv_3 - \kv_1$ & $\kv_{15} = 2 \kv_3 - \kv_2$ \\
      $\kv_{ 8} = 2 \kv_4 - \kv_1$ & $\kv_{12} = 2 \kv_4 - \kv_2$ & $\kv_{16} = 2 \kv_4 - \kv_3$ \\
      \hline
      \multicolumn{3}{c}{
        Remaining corners of tetrahedra }\\
      \multicolumn{3}{c}{
        that share surfaces with $T$ (blue balls in Fig. \ref{pointsfig}).} \\
      \hline
      \multicolumn{3}{c}{
        $\kv_{17} = \kv_4 - \kv_1 + \kv_2$ \hspace{1cm} $\kv_{18} = \kv_1 - \kv_2 + \kv_3$ }\\
      \multicolumn{3}{c}{
        $\kv_{19} = \kv_2 - \kv_3 + \kv_4$ \hspace{1cm} $\kv_{20} = \kv_3 - \kv_4 + \kv_1$ }
    \end{tabular}
  \end{ruledtabular}
\end{table}
We use a third order polynomial $A_T^3({\boldsymbol s})$ as $A_T^N({\boldsymbol s})$
in our implementation.
The sampling points used to construct $A_T^3({\boldsymbol s})$ are 
the corners of the tetrahedron $T$ (4 points) and the other 16 points given in
Table \ref{pointstable} and  Fig. \ref{pointsfig}.
As a result, $A_T^3({\boldsymbol s})$ becomes
\begin{align}
  A_T^3 &({\boldsymbol s}) 
  =  \frac{A_{1}}{2} u (u + 1) (2 - u) 
  +  \frac{A_{2}}{2} x (x + 1) (2 - x) \nonumber \\
  &+ \frac{A_{3}}{2} y (y + 1) (2 - y)
  +  \frac{A_{4}}{2} z (z + 1) (2 - z) \nonumber \\
  &- \frac{u^2 x}{6} (2 A_{ 5} + A_{14})
  -  \frac{x^2 y}{6} (2 A_{ 6} + A_{15}) \nonumber \\
  &- \frac{y^2 z}{6} (2 A_{ 7} + A_{16})
  -  \frac{z^2 u}{6} (2 A_{ 8} + A_{13}) \nonumber \\
  &- \frac{u^2 y}{6} (2 A_{ 9} + A_{11})
  -  \frac{x^2 z}{6} (2 A_{10} + A_{12}) \nonumber \\
  &- \frac{y^2 u}{6} (2 A_{11} + A_{ 9})
  -  \frac{z^2 x}{6} (2 A_{12} + A_{10}) \nonumber \\
  &- \frac{u^2 z}{6} (2 A_{13} + A_{ 8})
  -  \frac{x^2 u}{6} (2 A_{14} + A_{ 5}) \nonumber \\
  &- \frac{y^2 x}{6} (2 A_{15} + A_{ 6}) 
  -  \frac{z^2 y}{6} (2 A_{16} + A_{ 7}) \nonumber \\
  &+(A_{ 2} + A_{ 4}) x z (u + y)
  + (A_{ 1} + A_{ 3}) u y (x + z) \nonumber \\
  &- \left(A_{17} + \frac{A_{10} + A_{12}}{2} + \frac{A_{5} - A_{14}}{6} 
  + \frac{A_{13} - A_{ 8}}{6} \right) x z u \nonumber \\
  &- \left(A_{18} + \frac{A_{ 9} + A_{11}}{2} + \frac{A_{ 6} - A_{15}}{6} 
  + \frac{A_{14} - A_{ 5}}{6} \right) x y u \nonumber \\
  &- \left(A_{19} + \frac{A_{10} + A_{12}}{2} + \frac{A_{ 7} - A_{16}}{6}
  + \frac{A_{15} - A_{ 6}}{6} \right) x y z \nonumber \\
  &- \left(A_{20} + \frac{A_{ 9} + A_{11}}{2} + \frac{A_{ 8} - A_{13}}{6} 
  + \frac{A_{16} - A_{ 7}}{6} \right) y z u,
\end{align}
where $u=1-x-y-z$.
By substituting it into (\ref{eqn_lsm}), we obtain $A_T^{LSM}({\boldsymbol s})$: 
\begin{align}\label{eqn_barf}
  {\bar A}_{i} = \sum_{j = 1}^{20} P_{i j} A_{\kv_j},
\end{align}
where
\begin{align}\label{eqn_matrix}
  {\bf P} = ({\bf P}^{(1)}, {\bf P}^{(2)}, {\bf P}^{(3)}, {\bf P}^{(4)}, {\bf P}^{(5)}),
\end{align}
\begin{align}
  {\bf P}^{(1)} &= \frac{1}{1260} \begin{pmatrix}
    1440 & 0 & 30 & 0 \\
    0 & 1440 & 0 & 30 \\
    30 & 0 & 1440 & 0 \\
    0 & 30 & 0 & 1440 
  \end{pmatrix},
  \\
  {\bf P}^{(2)} &= \frac{1}{1260} \begin{pmatrix}
    -38 & 7 & 17 & -28 \\
    -28 & -38 & 7 & 17 \\
    17 & -28 & -38 & 7 \\
    7 & 17 & -28 & -38 
  \end{pmatrix} = {}^{t}{\bf P}^{(4)},
  \\
  {\bf P}^{(3)} &= \frac{1}{1260} \begin{pmatrix}
    -56 & 9 & -46 & 9 \\
    9 & -56 & 9 & -46 \\
    -46 & 9 & -56 & 9 \\
    9 & -46 & 9 & -56 
  \end{pmatrix}, 
  \\
  {\bf P}^{(5)} &= \frac{1}{1260} \begin{pmatrix}
    -18 & -18 & 12 & -18 \\
    -18 & -18 & -18 & 12 \\
    12 & -18 & -18 & -18 \\ 
    -18 & 12 & -18 & -18
  \end{pmatrix}.
\end{align}
We go through the same procedure for the orbital energy $\epsilon_{\kv}$.

We can consider this procedure in a different way;
when we calculate the contribution from a tetrahedron,
we use the linear tetrahedron method after
we have replaced matrix elements and orbital energies with 
those given in (\ref{eqn_barf}).
Using this idea, we represent the integration (\ref{k_int}) as
\begin{align}
  \int d^3 k A_{\kv} B(\epsilon_{\kv})
  = \sum_{\kv} A_{\kv} w_{\kv},
\end{align}
where $w_{\kv}$ is calculated as follows:
\begin{enumerate}
  \item We divide the BZ into tetrahedra.
  \item We calculate effective orbital energies as
    \begin{align}\label{eff_energy}
      {\bar \epsilon}_{i} = \sum_{j = 1}^{20} P_{i j} \epsilon_{\kv_j^T}
    \end{align}
    for the corners of each tetrahedron.
  \item We calculate the effective weight ${\bar w}^T_{i}$ 
    using the standard linear tetrahedron method with the effective orbital energy
    (\ref{eff_energy}). 
  \item $w_{\kv}$ is calculated as 
   \begin{align}
      w_{\kv} = \sum_T \sum_{i = 1}^{4} \sum_{j = 1}^{20} P_{i j} {\bar w}^T_{i} \delta(\kv -  \kv^T_j).
   \end{align}
\end{enumerate}
%
%
\section{Comparison with other integration schemes for actual calculations}
We implement our method in 
an {\it ab initio} electronic structure calculation code 
{\sc Quantum ESPRESSO}\cite{QE-2009}
which uses plane waves to represent Kohn-Sham (KS) orbitals.
Then, we test the effectiveness of the method
through calculations of phonons in two systems,
MgB$_2$ \cite{ISI:000167194300040} and 
fcc lithium at a high pressure (20 GPa),
based on DFPT
\cite{RevModPhys.73.515}(Appendix \ref{app_dfpt}).

\begin{figure}[!tb]
  \includegraphics[width=8.5cm]{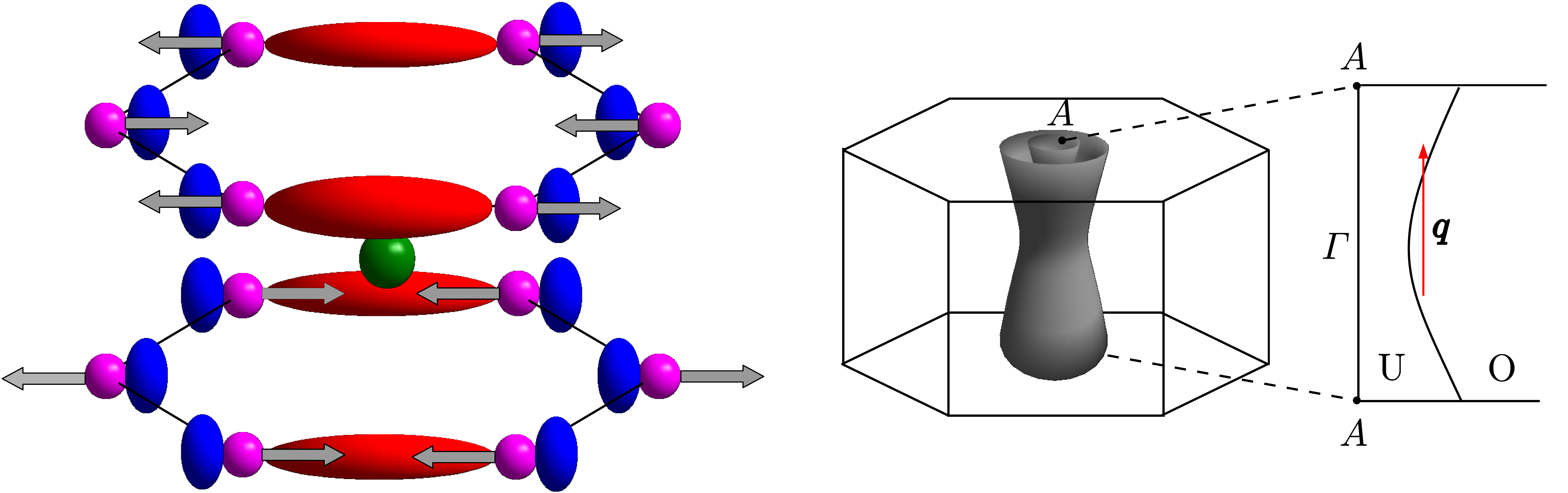}
  \caption{\label{mgb2fig}
    (left) The Mg centered Wigner-Seits cell of MgB$_2$.
    Green and purple spheres indicate Mg and B atoms respectively.
    The $\sigma$ orbital
    (blue and red isosurfaces of opposite signs.) 
    and the displacement pattern of 
    the intra-layer vibrational mode of the B atoms 
    with wave number $\qv$ at the $A$ point (arrows) are also depicted.
    (right) Schematic illustration of the Fermi surfaces of the $\sigma$ bands;
    the red arrow indicates the momentum vector of a phonon at the $A$ point
    which connects occupied (O) and unoccupied (U) regions in the vicinity of the Fermi surface.
 }
\end{figure}

%
%
Magnesium diboride has the highest $T_C$ (about 40 K)
out of the known phonon-type superconductors.
Many {\it ab initio} studies have been performed since it was discovered
\cite{PhysRevB.82.165111,PhysRevB.64.020501,
PhysRevLett.86.5771,PhysRevB.66.020513,PhysRevB.78.045124},
revealing that the high $T_C$ is a result of the strong interaction between
intra-layer vibrations of B atoms
and their covalent bonding orbitals ($\sigma$ bands) (Fig. \ref{mgb2fig}).
This strong coupling also softens phonon frequencies 
due to the screening of the ion-ion interaction;
this screening occurs due to linear responses of $\sigma$ electrons 
in the vicinity of the Fermi surfaces.
We have to evaluate these responses accurately to determine the phonon frequencies precisely.
%
%
Lithium exhibits a monatomic fcc structure at pressures between 7.5 and 39 GPa 
\cite{ISI:000165180400037}.
In this phase it becomes a superconductor. 
Its $T_C$ increases with pressure up to 30 GPa
\cite{PhysRevLett.91.167001, ISI:000179080400043, ISI:000178483100039}
because of the growth of the electron-phonon interaction.
The lower transverse acoustic mode at $\qv \approx \overrightarrow{\Gamma K}$ 
couples with electrons most strongly
in this material \cite{PhysRevB.82.184509}.
In this test, we consider the phonons of fcc Li at a pressure of 20 GPa.

We use norm-conserving pseudopotentials \cite{PhysRevLett.43.1494}
in calculations of MgB$_2$;
the cutoff energy of plane waves is set to 50 Ry.
In the calculations of fcc lithium, we use an ultrasoft pseudopotential\cite{PhysRevB.41.7892}.
We treat the electrons in the 1s orbitals as valence electrons\cite{PseudoLib}
and employ a cutoff energy of 80 Ry.
In both of these applications,
we use the GGA-PBE functional \cite{PhysRevLett.78.1396} and
the first-order Hermite-Gaussian function  \cite{PhysRevB.51.6773, PhysRevB.40.3616}
for broadening. 
%
%
\begin{figure}[!tb]
  \includegraphics[width=8.5cm]{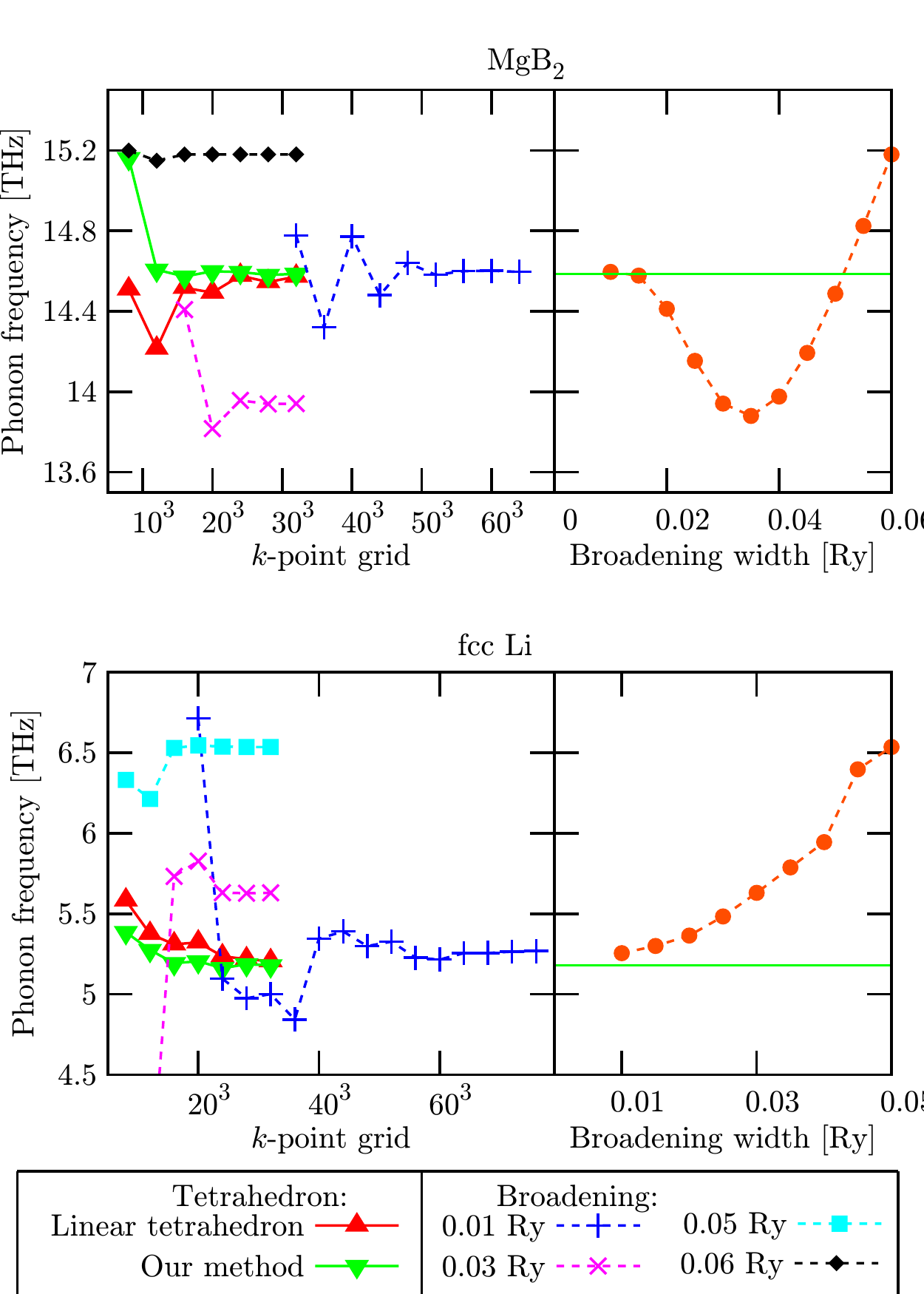}
  \caption{\label{freq}
    (left) The $\kv$ convergences of the frequencies 
    of the intra-layer vibrational mode of the B atoms at the $A$ point 
    in the BZ for MgB$_2$ (top)
    and
    the lower transverse acoustic mode at the $K$ point in the BZ 
    for fcc Li at 20 GPa (bottom)
    with a different $\kv$ integration method.
    $\blacktriangle$ and $\blacktriangledown$ with 
    red and green solid lines are 
    the results of the linear and improved tetrahedron methods;
    $+$, $\times$, $\blacksquare$, and $\blacklozenge$  with
    blue, purple, cyan, and black dashed lines
    denote the results of the broadening method 
    for widths of 0.01, 0.03, 0.05, and 0.06 Ry, respectively.
    Lines are guides for the eyes.
    (right) The frequency of these modes converged about the number of $\kv$
    at each broadening width ($\bullet$ with orange line);
    the green solid lines indicate the converged value obtained by our method.
  }
\end{figure}
We apply our method to the calculation of the frequency of the intra-layer vibrational mode of B
atoms at the $A$ point in the BZ (Fig. \ref{freq} top left).
The result of the improved tetrahedron method converges faster than that of
the linear tetrahedron method;
it converges with approximately $12^3$ $\kv$ points.
If we use a broadening method with a small broadening width (0.01 Ry), the result converges at
an unrealistically large number of $\kv$ points (about $50^3$ $\kv$ points).
On the other hand, 
using large broadening widths (0.03 Ry and 0.06 Ry),
convergence occurs at a lower number of $\kv$ points.
However, results are far away from the one converged about the broadening width;  
The complicated dependence of the convergence on the broadening width is shown 
in the top-right panel of Fig. \ref{freq}.
The result cannot be represented by a simple function, 
so it is difficult to extrapolate to a broadening width
of zero.

The bottom left panel of Fig. \ref{freq} shows the $\kv$ convergence 
of the lower transverse acoustic mode at the $K$ point in the BZ
for fcc Li at 20 GPa
calculated with the different integration schemes.
Our method achieves convergence very quickly;
it requires only $16^3$ $\kv$ points.
In this system, the result of the broadening method is 
very sensitive to the broadening width;
the error due to broadening is more than 25 \% at a width of 0.05 Ry;
hence, the broadening method is not suitable for this calculation.

%
%
We will show how the accuracy of the phonon calculations affects
the prediction of the superconducting transition temperature
within the framework of the following McMillan formula
\cite{PhysRev.167.331,Dynes1972615}:
\begin{align}
  \label{mcmillan}
  T_C = \frac{\omega_{\log}}{1.2} \exp \left(\frac{-1.04 (1+\lambda)}
  {\lambda - \mu^{*}(1+0.62 \lambda)}\right),
\end{align}
Here,
\begin{align}
  \label{lambda}
  &\lambda = \sum_{\qv \nu \kv n n'}  \frac{2}{D(\epsilon_{\rm F}) \omega_{\qv \nu}}
  \nonumber \\
  & \hspace{3em}
  \times |g_{n \kv n' \kv + \qv}^{\qv \nu}|^2 
  \delta(\epsilon_{n \kv}-\epsilon_{\rm F})
  \delta(\epsilon_{n' \kv+\qv}-\epsilon_{\rm F})
\end{align}
and
\begin{align}
  \label{omegalog}
  &\log(\omega_{\log}) = \frac{1}{\lambda} \sum_{\qv \nu \kv n n'} 
  \frac{2}{D(\epsilon_{\rm F}) \omega_{\qv \nu}} \log(\omega_{\qv \nu})
  \nonumber \\
  & \hspace{3em}
  \times |g_{n \kv n' \kv + \qv}^{\qv \nu}|^2 
  \delta(\epsilon_{n \kv}-\epsilon_{\rm F})
  \delta(\epsilon_{n' \kv+\qv}-\epsilon_{\rm F}),
\end{align}
where 
$\omega_{\qv \nu}$ is the phonon frequency with the wave number $\qv$ and the branch $\nu$,
$\epsilon_{n \kv}$ is the KS eigenvalue with the wave number $\kv$ and the band index $n$,
and $D(\epsilon_{\rm F})$ is the density of states per spin at the Fermi energy.
The electron-phonon coupling constant $g_{n \kv n' \kv + \qv}^{\qv \nu}$ is written in the form
\begin{align}
  g_{n \kv n' \kv + \qv}^{\qv \nu} = \sum_{\tau \alpha} 
  \frac{(\eta_{\qv \nu})_{\tau \alpha}}{\sqrt{M_{\tau} \omega_{\qv \nu}}}
  \Braket{n', \kv + \qv | 
    \frac{\delta v_{S}}{\delta R_{\tau \alpha}(\qv)}
    | n, \kv},
\end{align}
where $M_{\tau}$ is a mass of an ion,
$(\eta_{\qv \nu})_{\tau \alpha}$ is the unit displacement pattern of the phonon $(\qv, \nu)$,
$\Ket{n, \kv}$ is the KS orbital, and
${\delta v_{S}}/{\delta R_{\tau \alpha}(\qv)}$ is the linear response of the KS potential
with respect to the distortion of the wave number $\qv$;
$\tau$ and $\alpha$ are indices of an ion in the unit cell
and a direction in the Cartesian coordinate, respectively.
Although there are more precise methods to calculate $T_C$ such as
density functional theory for superconductors
\cite{PhysRevLett.60.2430, PhysRevB.72.024545},
we use this simple formula because we are only interested in changes in the results 
due to the $\kv$ integration
in the phonon calculations.

To evaluate the integrals in (\ref{lambda}) and (\ref{omegalog})~,
we use the linear tetrahedron method with
a $\qv$ grid of $6 \times 6 \times 4$ ($8 \times 8 \times 8$) and 
a $\kv$ grid of $24 \times 24 \times 18$ ($32 \times 32 \times 32$)
for MgB$_2$ (fcc Li), respectively.
Additionally, we calculate each $\omega_{\qv \nu}$ and
${\delta v_{S}}/{\delta R_{\tau \alpha}(\qv)}$ with different $\kv$ grids and
different $\kv$ integration schemes.
%
%
\begin{figure}[!tb]
  \includegraphics[width=8.5cm]{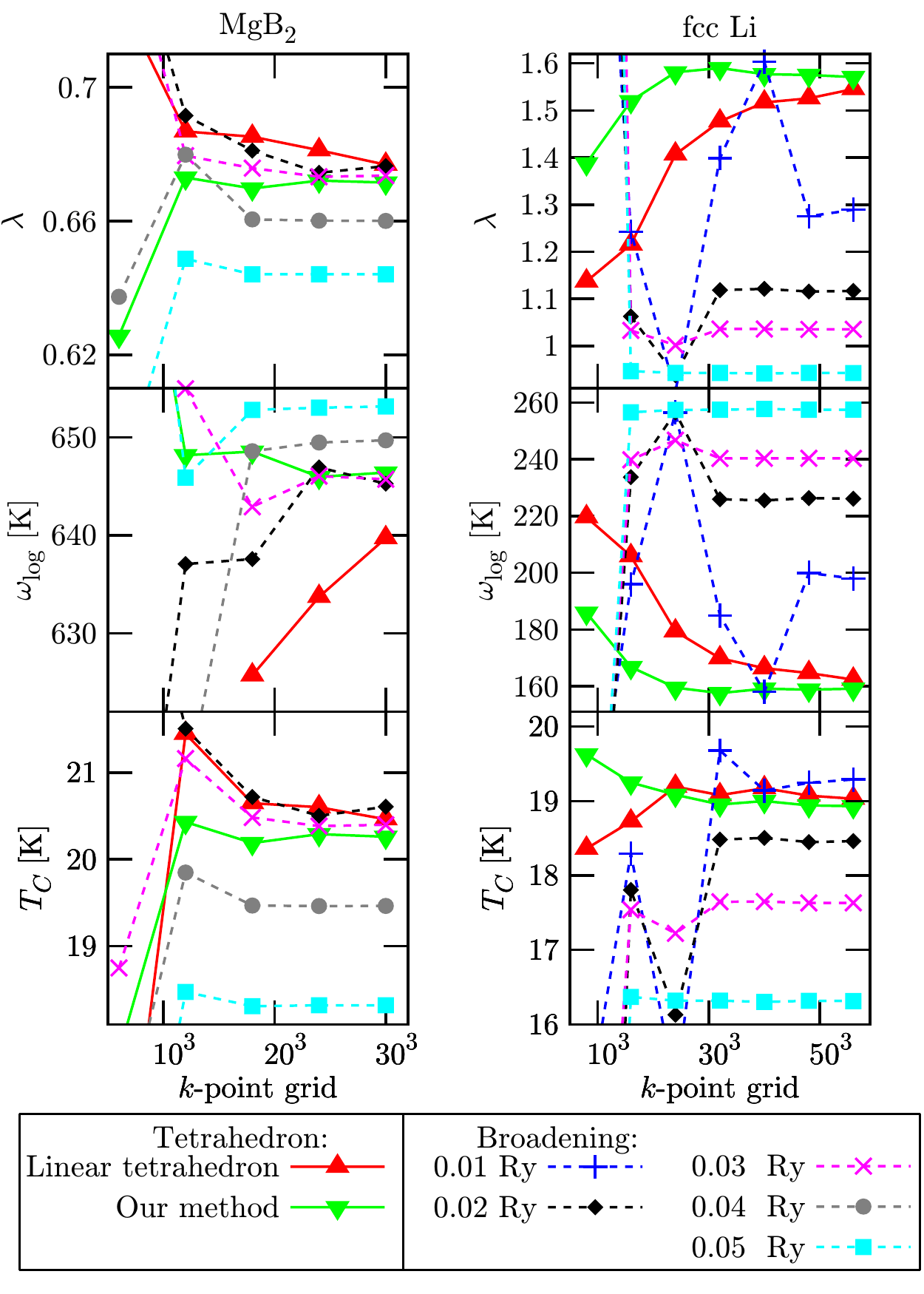}
  \caption{\label{lambdafig}
    The $\kv$ convergences of
    $\lambda$ (top), $\omega_{\log}$ (middle), and $T_C$ from the McMillan's formula (bottom)
    of MgB$_2$ (left) and fcc Li (right)
    calculated using $\omega_{\qv \nu}$ and 
    ${\delta v_{S}}/{\delta R_{\tau \alpha}(\qv)}$ with different $\kv$ integration schemes;
    $\blacktriangle$ and $\blacktriangledown$ with 
    red and green solid lines are 
    the results of the linear and improved tetrahedron methods;
    $\blacklozenge$,  $+$, $\times$, $\blacksquare$, and $\bullet$  with
    gray, blue, purple, black, and cyan dashed lines 
    denote the results of broadening methods 
    of widths 0.01, 0.02, 0.03, 0.04, and 0.05 Ry respectively;
    Lines are guides for the eyes. 
 }
\end{figure}

Figure \ref{lambdafig} shows the result of $\lambda$, $\omega_{\log}$, and
$T_C$ from the McMillan's formula ($\mu^{*} = 0.1$);
in both the MgB$_2$ and Li cases, we obtain very fast $\kv$ convergence 
using our method.
Comparing the $\kv$ converged result of our method to that of the broadening method with a width of 0.05 Ry,
we can see a large overestimate of the phonon frequencies occurs when the broadening method is used, resulting in 
an underestimated $\lambda$ and an overestimated $\omega_{\log}$.
Moreover, speeds of convergences about the broadening width 
for calculations of the $\lambda$ and $\omega_{\log}$ are very slow;
these results have not reach the convergence 
even for the broadening width of 0.01 Ry;
if we use smaller broadening width (such as 0.005 Ry),
we need an unrealistic number of $\kv$ points to obtain the $\kv$-converged result.
%
%
\section{Conclusion}
We introduced an improvement to the tetrahedron method based on
the third order interpolation and the least square method
that reduces the number of $\kv$ points required to obtain converged results of
the BZ integrations.
Our method is applicable to various kinds of $\kv$-integration;
in particular, it is efficient for calculations of phonons and response functions
because the associated computational costs are large
and the Bl\"ochl correction is not applicable to these calculations.
We demonstrated this effectiveness 
through calculations of phonon frequencies in MgB$_2$ and fcc Li.
%
%
\begin{acknowledgments}
This work was supported by 
the Elements Strategy Initiative Center for Magnetic Materials (ESICMM)
under the outsourcing project of MEXT.
The numerical calculations were performed using Fujitsu FX10s
at the Information Technology Center 
and the Institute for Solid State Physics, The University of Tokyo.
\end{acknowledgments}
%
%
%
\appendix
\section{Calculation of weights for DFPT}\label{app_dfpt}
\begin{figure}[!tb]
  \includegraphics[width=8.5cm]{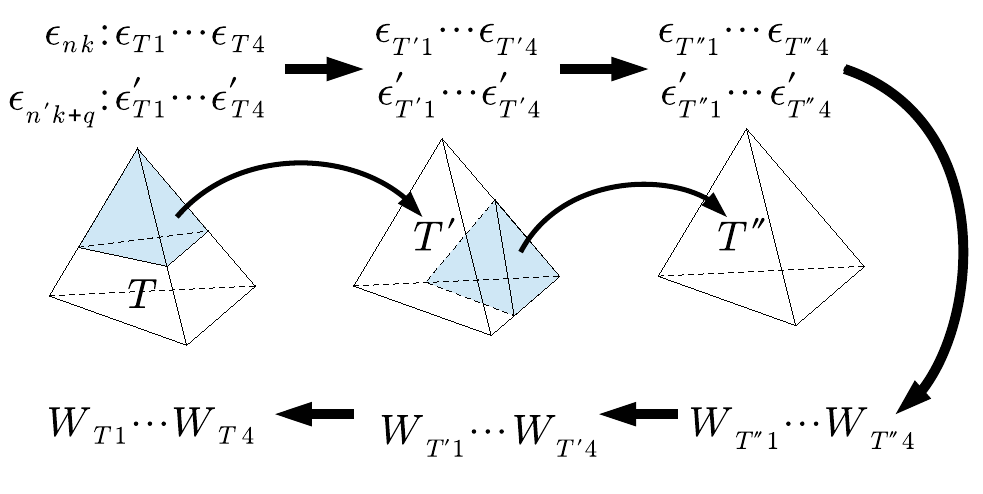}
  \caption{\label{fllowfig}Flow of the calculation of weights.
    We divide the tetrahedra two times to
    cut out regions where two Heaviside functions become one.}
\end{figure}

%
%
The integration weights for the DFPT calculations of phonon frequencies are 
different from those of the total energy, $\theta(\epsilon_{\rm F}-\epsilon_{n k})$,
or the density of states, $\delta(\epsilon - \epsilon_{n k})$.
They are 
\begin{align}
  W^{(1)}_{n n' k} &= \frac{
    \theta(\epsilon_{n' k+q} - \epsilon_{\rm F})\theta(\epsilon_{\rm F} - \epsilon_{n k}) 
  }{\epsilon_{n k} - \epsilon_{n' k+q}}
  \\
  W^{(2)}_{n n' k} &= \theta(\epsilon_{\rm F} - \epsilon_{n k}) 
  \theta(\epsilon_{n k} - \epsilon_{n' k+q}).
\end{align}
In integrations with weights that contain products of two step functions, 
only regions where both Heaviside functions become 1 contribute to the results;
therefore, we divide the tetrahedra two times to cut out these regions (Fig. \ref{fllowfig}).
We will explain how to calculate $W^{(1)}_{n n' k}$.
\begin{enumerate}
\item We divide a sub-cell into six tetrahedra.
\item We cut out one or three tetrahedra $T'$ where
  $\theta(\epsilon_{\rm F} - \epsilon_{n k})=1$ from tetrahedron $T$
  and evaluate $\epsilon_{n k},\epsilon_{n' k+q}$ at the corners of $T''$ as
  \begin{align}
    \epsilon_{T' i} = \sum_{j=1}^4 F_{i j}(
    \epsilon_{\rm F}-\epsilon_{T 1}, \cdots, \epsilon_{\rm F}-\epsilon_{T 4}) 
    \epsilon_{T j},
  \end{align}
  through linear interpolation (Appendix \ref{app_tetradevide}).
  Here
  $\epsilon_{T 1}, \cdots \epsilon_{T 4}$ and $\epsilon'_{T 1}, \cdots \epsilon'_{T 4}$ 
  are $\epsilon_{n k}$ and $\epsilon_{n' k+q}$, respectively, on the corners of $T$,
  where $\epsilon_{T 1} \leq \epsilon_{T 2} \leq \epsilon_{T 3} \leq \epsilon_{T 4}$.
\item We cut out one or three tetrahedra $T''$ where
  $\theta(\epsilon_{n' k+q} - \epsilon_{\rm F})=1$ from tetrahedron $T'$.
  The orbital energies are calculated as 
  \begin{align}
    \epsilon_{T'' i} = \sum_{j=1}^4 F_{i j}(
    \epsilon'_{T' 1} - \epsilon_{\rm F}, \cdots, \epsilon'_{T' 4} - \epsilon_{\rm F}) 
    \epsilon_{T' j}.
  \end{align}
\item The weights of the corners of $T''$ are
  (Appendix \ref{app_lindhard})
  \begin{align}\label{eqn_lindhard}
    W_{T'' i}^{(1)} &= -V_{T''} \sum_{j=1,j \ne i}^4
    \frac{d_j^2
      \left( \frac{\ln d_j - \ln d_i}{d_j - d_i} d_j - 1 \right)}
         {\prod_{k=1,k \ne j}^4 (d_j - d_k)},
  \end{align}
  where $d_i=\epsilon'_{T'' i}-\epsilon_{T'' i}$.
\item We calculate the weights of the corners of $T'$ from those of $T''$.
  \begin{align}
    W_{T' i}^{(1)} = \sum_{j=1}^4 F_{j i}(
    \epsilon'_{T' 1} - \epsilon_{\rm F}, \cdots, \epsilon'_{T' 4} - \epsilon_{\rm F}) 
    W_{T'' j}^{(1)}.
  \end{align}
\item We calculate the weights of the corners of $T$ from those of $T'$.
  \begin{align}
    W_{T i}^{(1)} = \sum_{j=1}^4 F_{j i}(
    \epsilon_{\rm F}-\epsilon_{T 1}, \cdots, \epsilon_{\rm F}-\epsilon_{T 4}) 
    W_{T' j}^{(1)}.
  \end{align}
\item Finally, we sum up the contributions from all tetrahedra.
  \begin{align}
    W_{n n' k}^{(1)} = \sum_{T=1}^{6 N_{\kv}} \sum_{i=1}^4 W_{T i}.
  \end{align}
\end{enumerate}
\section{How to divide a tetrahedron}\label{app_tetradevide}
%
%
\begin{figure}[!tb]
  \includegraphics[width=8.5cm]{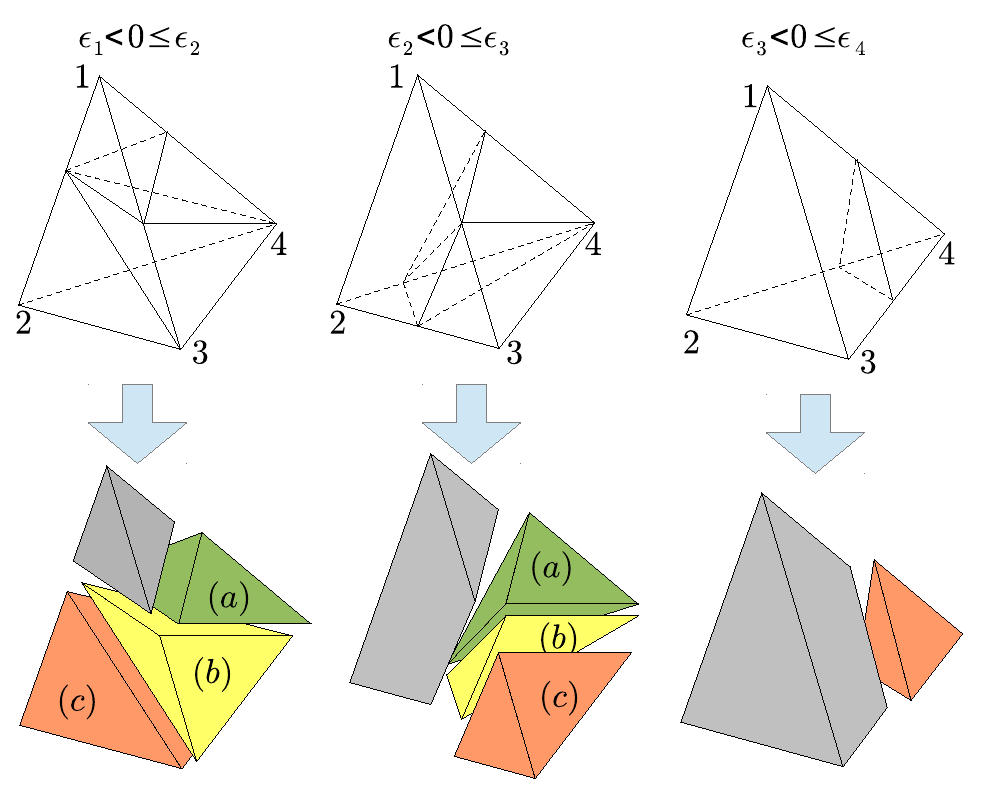}
  \caption{\label{dividefig}How to divide a tetrahedron 
    in the case of $\epsilon_1 \leq 0 < \epsilon_2$ (left), 
    $\epsilon_2 \leq 0 < \epsilon_3$ (center), and
    $\epsilon_3 \leq 0 < \epsilon_4$ (right).} 
\end{figure}
We will explain how to cut out tetrahedra $T'$ in the case of $\theta(\epsilon)=1$
from tetrahedron $T$.
We represent $\epsilon_{n k}$ at the corners of $T$ as $\epsilon_1, \cdots, \epsilon_4$,
where $\epsilon_1 \leq \epsilon_2 \leq \epsilon_3 \leq \epsilon_4$.
We define $a_{i j}=-\epsilon_j/(\epsilon_i - \epsilon_j)$.
In all cases
\begin{align}
  V_{T'} = V_{T} \left| \det \left( F \right) \right|.
\end{align} 
\begin{enumerate}
\item $0 \leq \epsilon_1$: \\
  We cut out no tetrahedra 
  because $\theta(\epsilon)$ becomes 1 over the entire tetrahedron in this case.
  \begin{align}
    F &= 
    \begin{pmatrix}
      1 & 0 & 0 & 0 \\
      0 & 1 & 0 & 0 \\
      0 & 0 & 1 & 0 \\
      0 & 0 & 0 & 1
    \end{pmatrix}
  \end{align}
\item $\epsilon_1 \leq 0 < \epsilon_2$: \\
  Three tetrahedra are cut out (Fig. \ref{dividefig} left).
  \begin{align}
    (a) \qquad F &= 
    \begin{pmatrix}
      a_{1 2} & a_{2 1} & 0       & 0 \\
      a_{1 3} & 0       & a_{3 1} & 0 \\
      a_{1 4} & 0       & 0      & a_{4 1} \\
      0      & 0       & 0       & 1
    \end{pmatrix}
    \\
    (b) \qquad F &= 
    \begin{pmatrix}
      a_{1 2} & a_{2 1} & 0       & 0 \\
      a_{1 3} & 0       & a_{3 1} & 0 \\
      0      & 0       & 1       & 0 \\
      0      & 0       & 0       & 1
    \end{pmatrix}
    \\
    (c) \qquad F &= 
    \begin{pmatrix}
      a_{1 2} & a_{2 1} & 0 & 0 \\
      0      & 1       & 0 & 0 \\
      0      & 0       & 1 & 0 \\
      0      & 0       & 0 & 1
    \end{pmatrix}
  \end{align}
\item $\epsilon_2 \leq 0 < \epsilon_3$: \\
  Three tetrahedra are cut out (Fig. \ref{dividefig} center).
  \begin{align}
    (a) \qquad F &= 
    \begin{pmatrix}
      a_{1 3} & 0       & a_{3 1} & 0      \\
      a_{1 4} & 0       & 0      & a_{4 1} \\
      0      & a_{2 4}  & 0      & a_{4 2} \\
      0      & 0       & 0      & 1
    \end{pmatrix}
    \\
    (b) \qquad F &= 
    \begin{pmatrix}
      a_{1 3} & 0       & a_{3 1} & 0      \\
      0      & a_{2 3} & a_{3 2} & 0      \\
      0      & a_{2 4}  & 0      & a_{4 2} \\
      0      & 0       & 0       & 1
    \end{pmatrix}
    \\
    (c) \qquad F &= 
    \begin{pmatrix}
      a_{1 3} & 0       & a_{3 1} & 0 \\
      0      & a_{2 3}  & a_{3 2} & 0 \\
      0      & 0       & 1      & 0  \\
      0      & 0       & 0      & 1
    \end{pmatrix}
  \end{align}
\item $\epsilon_3 \leq 0 < \epsilon_4$: \\
  One tetrahedron is cut out (Fig. \ref{dividefig} right).
  \begin{align}
    F &= 
    \begin{pmatrix}
      a_{1 4} & 0       & 0      & a_{4 1} \\
      0      & a_{2 4}  & 0      & a_{4 2} \\
      0      & 0       & a_{3 4} & a_{4 3} \\
      0      & 0       & 0       & 1
    \end{pmatrix}
  \end{align}
\end{enumerate}
%
%
\section{Calculation of $W^{(1)}_{T''}$}\label{app_lindhard}
We represent the matrix elements at the corners of the tetrahedron as $f_1,\cdots,f_4$.
We evaluate the integral
\begin{align}
  \Braket{\frac{A}{d}}_{T''} = \int_{T''} d^3 k \frac{A_{\kv}}{d_{\kv}} 
\end{align}
using linear interpolation to obtain
\begin{align}
  A_{T''} \approx &6 V'' \int^1_0 dx \int_0^{1-x} dy \int_0^{1-x-y} dz 
  \nonumber \\
  & \times \frac{A_1 + (A_2 - A_1)x + (A_3 - A_1)y + (A_4 - A_1)z }
            {d_1 + (d_2 - d_1)x + (d_3 - d_1)y + (d_4 - d_1)z }
  \nonumber \\
  \equiv &\sum_{i=1}^4 A_i W_{T'' i},
\end{align}
where
\begin{align}
  W_{T'' i} = 6 V'' \int_0^1 &dx_1 \int^1_0 dx_2 \int_0^1 dx_3 \int_0^1 dx_4 
  \nonumber \\
  &\times \frac{x_i \delta(x_1+x_2+x_3+x_4-1)}{d_1 x_1 + d_2 x_2 + d_3 x_3 + d_4 x_4}.
\end{align}
This in turn yields (\ref{eqn_lindhard}).

\bibliography{kawamura}

\end{document}